\documentclass[11pt]{article}

\newsavebox{\foobox}
\newcommand{\slantbox}[2][0]{\mbox{%
        \sbox{\foobox}{#2}%
        \hskip\wd\foobox
        \pdfsave
        \pdfsetmatrix{1 0 #1 1}%
        \llap{\usebox{\foobox}}%
        \pdfrestore
}}
\newcommand\unslant[2][-.25]{\slantbox[#1]{$#2$}}

\newcommand{\mpi}{\text{\unslant[-.18]\pi}}
\newcommand{\mdelta}{\text{\unslant[-.18]\delta}}

\usepackage[left=2cm, right=2cm, top=2.5cm, bottom=2.5cm]{geometry}
\usepackage[x11names]{xcolor}
\usepackage{fancyhdr, amssymb, cancel, amsmath, graphicx, pgfplots, tikz}
\usepackage{isomath}

\usetikzlibrary{shadows}

\newcommand{\stylecolor}{violet}

\usepackage[labelfont={bf,sf, color=\stylecolor}, margin={1.5cm,0cm}]{caption}

\usepackage[colorlinks=true, urlcolor=\stylecolor!70!white, linkcolor=\stylecolor, citecolor=\stylecolor!70!white, hyperindex=true, linktocpage=true]{hyperref}

\usepackage[explicit]{titlesec}

\newcommand*\sectionlabel{}
\titleformat{\section}
  {\gdef\sectionlabel{}
   \Large\bfseries\scshape}
  {\gdef\sectionlabel{\thesection }}{0pt}
  {\begin{tikzpicture}[remember picture,overlay]
       \end{tikzpicture}
  }
\titlespacing*{\section}{0pt}{0pt}{0pt}

\newcommand*\subsectionlabel{}
\titleformat{\subsection}
  {\gdef\subsectionlabel{}
   \large\bfseries\scshape}
  {\gdef\subsectionlabel{\thesubsection  }}{0pt}
  {\begin{tikzpicture}[remember picture]
    	\draw (-0.15, 0) node[left] {\color{\stylecolor} \textsf{\subsectionlabel}};
	\draw (0.15, 0) node[right] {\color{\stylecolor} \textsf{#1}};
	\fill[color=\stylecolor] (-0.05, -0.23) rectangle (0.05, 0.23);
       \end{tikzpicture}
  }
\titlespacing*{\subsection}{-4pt}{10pt}{0pt}

\newcommand*\subsubsectionlabel{}
\titleformat{\subsubsection}
  {\gdef\subsubsectionlabel{}
   \bfseries\scshape}
  {\gdef\subsubsectionlabel{\thesubsubsection.\ \  }}{0pt}
  {\begin{tikzpicture}[remember picture]
    	\draw (0, 0) node[left] {\color{\stylecolor} \textsf{\subsubsectionlabel}};
	\draw (0, 0) node[right] {\color{\stylecolor} \textsf{#1}};
       \end{tikzpicture}
  }
\titlespacing*{\subsubsection}{-4pt}{7pt}{0pt}

\pgfplotsset{every axis legend/.append style={at={(1.02,1)},anchor=north west}}

\begin{document}

\allowdisplaybreaks

\pagestyle{fancy}
\renewcommand{\headrulewidth}{0pt}
\fancyhead{}

\fancyfoot{}
\fancyfoot[C] {\textsf{\textbf{\thepage}}}

\begin{equation*}
\begin{tikzpicture}
\draw (\textwidth, 0) node[text width = \textwidth, right] {\color{white} easter egg};
\end{tikzpicture}
\end{equation*}

\begin{equation*}
\begin{tikzpicture}
\draw (0.5\textwidth, -3) node[text width = \textwidth] {\huge  \textsf{\textbf{Breakdown of emergent Lifshitz symmetry in  \\ \vspace{0.07in} holographic matter with  Harris-marginal disorder}} };
\end{tikzpicture}
\end{equation*}
\begin{equation*}
\begin{tikzpicture}
\draw (0.5\textwidth, 0.1) node[text width=\textwidth] {\large \color{black}  \textsf{Koushik Ganesan and Andrew Lucas}};
\draw (0.5\textwidth, -0.5) node[text width=\textwidth] {\small\textsf{Department of Physics and Center for Theory of Quantum Matter, University of Colorado, Boulder, CO 80309, USA}};
\end{tikzpicture}
\end{equation*}
\begin{equation*}
\begin{tikzpicture}
\draw (0, -13.1) node[right, text width=0.75\paperwidth] {\texttt{koushik.ganesan@colorado.edu, andrew.j.lucas@colorado.edu}};
\draw (\textwidth, -13.1) node[left] {\textsf{\today}};
\end{tikzpicture}
\end{equation*}
\begin{equation*}
\begin{tikzpicture}
\draw[very thick, color=\stylecolor] (0.0\textwidth, -5.75) -- (0.99\textwidth, -5.75);
\draw (0.12\textwidth, -6.25) node[left] {\color{\stylecolor}  \textsf{\textbf{Abstract:}}};
\draw (0.53\textwidth, -6) node[below, text width=0.8\textwidth, text justified] {\small We revisit the theory of strongly correlated quantum matter perturbed by Harris-marginal random-field disorder, using the simplest holographic model.  We argue that for weak disorder, the ground state of the theory is not Lifshitz invariant with a non-trivial disorder-dependent dynamical exponent, as previously found.  Instead, below a non-perturbatively small energy scale, we predict infrared physics becomes independent of the disorder strength.};
\end{tikzpicture}
\end{equation*}

\tableofcontents

\begin{equation*}
\begin{tikzpicture}
\draw[very thick, color=\stylecolor] (0.0\textwidth, -5.75) -- (0.99\textwidth, -5.75);
\end{tikzpicture}
\end{equation*}

\titleformat{\section}
  {\gdef\sectionlabel{}
   \Large\bfseries\scshape}
  {\gdef\sectionlabel{\thesection }}{0pt}
  {\begin{tikzpicture}[remember picture]
	\draw (0.2, 0) node[right] {\color{\stylecolor} \textsf{#1}};
	\draw (0.0, 0) node[left, fill=\stylecolor,minimum height=0.27in, minimum width=0.27in] {\color{white} \textsf{\sectionlabel}};
       \end{tikzpicture}
  }
\titlespacing*{\section}{0pt}{20pt}{5pt}

\section{Introduction}
The past decade has seen a flurry of activity on the applications of gauge-gravity duality (or holography) to condensed matter physics \cite{review1, review2, koenraadbook, mybook}.  In a nutshell, holography maps the dynamics of a strongly correlated quantum field theory (with a large $N$ microscopic number of degrees of freedom at each point in spacetime) to a dual problem in classical gravity in an asymptotically anti de Sitter (AdS) spacetime.  This duality maps problems which can be completely intractable in field theory into problems about classical gravity, which are therefore more tractable.

This paper is about a particularly challenging problem:  the ground state of a strongly correlated system in the presence of disorder \cite{halperin, xgwen, fisher1, fisher2}.  Holography is a particularly appealing tool for these systems because the disorder can be treated at a quite direct level: it is simply dual in the bulk gravitational description to an inhomogeneous boundary condition on fields at the boundary of AdS.  In principle, we can solve the classical bulk equations \emph{before} disorder averaging, thus giving a more physical treatment of  the randomness \cite{hartnollimpure},\footnote{Holographic models allow us to study the physically correct prescription of \emph{quenched disorder}: one calculates for example, the disorder averaged free energy $\mathbb{E}[F]$ where $\mathbb{E}[\cdots]$ denotes the disorder average.  The easier computation to perform is over \emph{annealed disorder}, where one evaluates the average of the partition function $\mathbb{E}[e^{-\beta F}]$.   Generally in disordered systems, $\mathbb{E}[\mathrm{e}^{-\beta F}]$ and $\exp[-\beta \mathbb{E}[F]]$ are not the same, and may even suggest entirely different phase diagrams!  (Usually the annealed average is dominated by measure zero fluctuations which are enhanced in the average.) In many field theory analyses \cite{halperin, xgwen, fisher1, fisher2}, quenched disorder is treated using the replica method, which amounts to creating $n$ coupled ``replicas"  of the theory, evaluating the partition function for integer $n$, and taking the $n\rightarrow 0$ limit.   This is, in our view, rather unphysical, which is why holographic models which can directly evaluate quenched disorder averages are so valuable.} albeit in a somewhat mysterious dual description.  Indeed, this problem was addressed in a series of papers \cite{santosdisorder1, santosdisorder2, santosdisorder3} which found that in a variety of different holographic models, \emph{marginal} disorder leads to the emergence of a non-relativistic scaling symmetry called Lifshitz symmetry.  We will define these terms properly in the following section.   This emergent scale invariance was later reproduced on general field theoretic grounds \cite{aharony}.  Further holographic studies of similar systems include \cite{awpeet, garciagarcia, matteo}.  The Lifshitz scaling was also seen at weak coupling many decades ago \cite{cardy}.

The purpose of this paper is to revisit the arguments for Lifshitz invariance of the low energy theory.  We study the regime in which disorder is weak, and argue that although there is an emergent scale invariance over an extraordinarily large range of energy scales, non-perturbative effects ultimately destroy scale invariance and logarithmically slowly push the theory towards a logarithmically-modified ``almost conformal theory" of the true infrared.  Our conclusion is justified by a heuristic approximation for solving the holographic equations of motion in the bulk in the presence of disorder, which was first sketched in \cite{mybook}.   While our conclusions appear superficially in tension with the previous numerical results (which rely on fewer assumptions), the breakdown of Lifshitz scaling is a non-perturbative effect in the dimensionless disorder strength $\bar V$, which arises at energy scales $\Lambda \exp[-1/\bar V^2]$ (with $\Lambda$ a UV scale).   The analytical analyses in previous literature such as \cite{santosdisorder1} did not extend to energy scales parametrically beyond this deeply infrared scale. Furthermore, as our conclusions do not appear to be sensitive to details of the holographic model, and are hinted at by a simple and very general argument, our results  suggest interesting physics about non-perturbative physics of marginally disordered field theories more generally \cite{aharony}.

\section{Minimal holographic model of disorder}
Let us now introduce the model we study more carefully.  We assume the reader has familiarity with gauge-gravity duality: pedagogical texts on the subject include \cite{review1, review2, koenraadbook, mybook}.  We are interested in studying some strongly coupled $1+1$-dimensional conformal field theory (CFT) of action $S_0$, perturbed weakly by disorder that couples to a Lorentz scalar operator $\mathcal{O}$ of dimension $\Delta$: \begin{equation}
S = S_0 - \int \mathrm{d}t\mathrm{d}x \; h(x)\mathcal{O}(x).
\end{equation}
The classical background field $h(x)$ corresponds to random-field disorder.  Denoting the disorder average by $\mathbb{E}[\cdots]$, we choose $h(x)$ to be Gaussian disorder with mean and variance \begin{subequations}\begin{align}
\mathbb{E}[h(x)] &= 0 , \\ 
\mathbb{E}[h(x)h(y)] &= \bar V^2 \mdelta(x-y). \label{eq:Ehxhy}
\end{align}\end{subequations}
Note that $h(x)$ depends on space but not time. The disorder is defined to be weak when $\bar V^2$ is small; however we caution the reader that such weak disorder does not mean that pointwise fluctuations in $h(x)$ are small in a given realization (indeed (\ref{eq:Ehxhy}) says the average value of $h(x)^2$ is divergent, although this is an artifact of the continuum theory).   We are interested in theories where in the CFT this disorder is marginal: namely, $\bar V$ is dimensionless.  This occurs when $[h] = \frac{1}{2}$, or when \begin{equation}
\Delta = \frac{3}{2}. \label{eq:dimO}
\end{equation}  
Note that an ordinary scalar field would have been marginal when $\Delta=2$;  the discrepancy arises from the fact that the random coupling $h(x)$ does not have the same scaling dimension as a uniform coupling, due to the $\mdelta$-function in (\ref{eq:Ehxhy}) which is not dimensionless.  We also note that while $1+1$-dimensional CFTs are far more tractable than their higher-dimensional counterparts, within holography, we do not expect our conclusions to be sensitive to spacetime dimension.

We study this perturbed CFT holographically using a three dimensional gravity theory.  The bulk holographic action is \begin{equation}
S = \int \mathrm{d}^3x \sqrt{-g}\left(R + 2 - \frac{1}{2}\nabla_a \phi \nabla^a \phi - \frac{1}{2}m^2 \phi^2 \right) \label{eq:minimal}
\end{equation}
We have rescaled coordinates and fields such that the gravitational coupling constant and the AdS radius are 1, without loss of generality.  The dynamical bulk fields are the metric $g_{ab}$ and a real scalar field $\phi$; indices $a,b,\ldots$ denote bulk coordinates.  The real scalar field $\phi$ is dual to $\mathcal{O}$, and via the holographic dictionary, this requires 
\begin{equation}
m^2 = -\frac{3}{4}. \label{eq:m234}
\end{equation}

This holographic model was previously studied analytically and numerically in \cite{santosdisorder1};  both methods suggested that the infrared theory of this disordered CFT had Lifshitz scale invariance.  This means that the theory is scale invariant, but that time and space scale separately:  \begin{equation}
[t] = z[x],
\end{equation}
where $z$ is called the dynamical critical exponent.  In our model, and using our normalizations,  when $\bar V$ is small, \cite{mybook, santosdisorder1}
\begin{equation}
z = 1 + \frac{\bar V^2}{8} + \cdots.  \label{eq:zlifshitz}
\end{equation}

The emergence of Lifshitz scaling in this theory introduces a puzzle which, to our understanding, has not been remarked upon in the literature.  The generalization of (\ref{eq:dimO}) to a model in $d$ spatial dimensions (not including time), and with dynamical critical exponent $z$, is that disorder is marginal when \cite{mybook, lss, harris}  \begin{equation}
\Delta_{\mathrm{marginal}} = \frac{d}{2}+z. \label{eq:harrismarginal}
\end{equation} 
In the holographic dictionary, the mass $m$ of $\phi$ is related to its operator dimension as \begin{equation}
m^2 = \Delta(\Delta-d-z). \label{eq:mdelta}
\end{equation}
Plugging in $d=1$ and using the value of $m^2$ from (\ref{eq:m234}) and $z$ from (\ref{eq:zlifshitz}), we find that \begin{equation}
\Delta \approx \frac{3}{2} + \frac{3\bar V^2}{16}. \label{eq:deltalifshitz}
\end{equation}
Comparing (\ref{eq:harrismarginal}) and (\ref{eq:deltalifshitz}), we conclude that disorder becomes \emph{irrelevant}.  So how can this irrelevant disorder continue to support a Lifshitz-invariant bulk spacetime?

\section{Constructing the bulk geometry}
The coupled Einstein-scalar system reads \begin{subequations}\label{eq:generalsystem}\begin{align}
R_{ab} - \frac{R}{2}g_{ab} - g_{ab} &= \frac{1}{2}\left[\partial_a \phi \partial_b \phi - \frac{1}{2}g_{ab}\left(\partial_c\phi \partial^c\phi -\frac{3}{4}\phi^2\right)\right], \label{eq:einsteineq} \\
\nabla_a \nabla^a \phi &= m^2\phi. \label{eq:scalareom}
\end{align}\end{subequations}
Let $r$ denote the bulk radial coordinate, with $r=0$ corresponding to an asymptotically AdS boundary. (\ref{eq:generalsystem}) is to be solved subject to the following boundary conditions at $r=0$: \begin{subequations}\begin{align}
\phi(x,r\rightarrow 0) &= r^{1/2}\int \frac{\mathrm{d}k}{2\mpi} \mathrm{e}^{\mathrm{i}kx} h(k), \\
\mathrm{d}s^2(r\rightarrow 0) &= \frac{\mathrm{d}r^2+\mathrm{d}x^2-\mathrm{d}t^2}{r^2}.
\end{align}\end{subequations}
We assume that $h(k)$ are Gaussian random fields with zero mean, as before: $\mathbb{E}[h(k)]=0$.  However, for reasons that will shortly become clear, we will choose the variance of the disorder to be \begin{equation}
\mathbb{E}\left[h(k)h(q)\right] = 2\mpi  \bar V^2 \mathrm{\Theta}(\Lambda-|k|) \mdelta(k+q), \label{eq:momspacedis}
\end{equation}
where $\Lambda$ serves as an ultraviolet cutoff on the disorder.

Let us now (approximately) solve the bulk equations of motion, building off of ideas described in \cite{mybook}.  We make an ansatz that so long as $\bar V$ is small, the metric is well approximated by a homogeneous metric: \begin{equation}
\mathrm{d}s^2 \approx \frac{\mathrm{d}r^2}{r^2}+A(r)\mathrm{d}x^2 - B(r)\mathrm{d}t^2, \label{eq:homogeometry}
\end{equation}
where $r$ represents the bulk radial coordinate ($r=0$ corresponding to the asymptotically AdS boundary).  One justification for our assumption that the metric stays homogeneous is that inhomogeneity in the metric is $\mathrm{O}(\bar V^2)$ at finite wave number and thus feeds back to give $\mathrm{O}(\bar V^4)$ corrections to the homogeneous part of the metric, while on average the inhomogeneous contributions do not contribute to the metric.  Since the scalar field distorts the homogeneous metric at $\mathrm{O}(\bar V^2)$ we expect it is qualitatively more important (although this is a guess, not a proof). Anyway, the above metric ansatz fixes the gauge for Einstein's equations.  This ansatz will capture the same physics as observed in earlier work \cite{santosdisorder1, awpeet, garciagarcia}.  In order to neglect the spatial inhomogeneity in the metric introduced via the inhomogeneous scalar field, we will disorder average the right hand side of (\ref{eq:einsteineq}) before solving for $A(r)$ and $B(r)$.  We obtain the following two independent equations: \begin{subequations}\label{eq:system}\begin{align}
r\sqrt{\frac{A}{B}}\partial_r \left(r\sqrt{\frac{A}{B}}\partial_rB\right) &= 4A + \frac{3}{4}A\mathbb{E}\left[\phi^2\right], \\
\frac{1}{r}\sqrt{\frac{A}{B}}\partial_r \left(\frac{r\partial_rB}{\sqrt{AB}}\right) &= \frac{\mathbb{E}\left[(\partial_x\phi)^2\right]}{r^2A} - \mathbb{E}\left[(\partial_r\phi)^2\right],
\end{align}\end{subequations}
which give us a self-consistent ``mean-field" bulk geometry which is spatially homogeneous.   

It is useful to begin by simply treating $\bar V$ as a strict perturbative parameter.  In this case, we can explicitly carry out the first order in the perturbative expansion.  This was done in \cite{santosdisorder1}.   We find that at first order in $\bar V$, \begin{equation}
\phi(k,r) = h(k) \sqrt{r}\mathrm{e}^{-|k|r}.
\end{equation}
Since at leading order \begin{subequations}\begin{align}
\mathbb{E}\left[\phi^2\right] &= \bar V^2 \left(1-\mathrm{e}^{-2r\Lambda}\right) = \int\limits_{-\Lambda}^\Lambda \frac{\mathrm{d}k}{2\mpi} \times 2\mpi \bar V^2 r \mathrm{e}^{-2|k|r}  , \\
\mathbb{E}\left[(\partial_x\phi)^2 - (\partial_r\phi)^2\right] &= \frac{\bar V^2\left(1-\mathrm{e}^{-2r\Lambda}(1+4r\Lambda)\right)}{4r^2},
\end{align}\end{subequations}
we arrive at a homogeneous correction to the metric:
\begin{subequations}\begin{align}
A(r) &= \frac{1}{r^2} -   \frac{\bar V^2}{4r^2} \left(1-\mathrm{e}^{-2r\Lambda}\right), \\
B(r) &=  \frac{1}{r^2} + \frac{\bar V^2}{4r^2} \left(\mathrm{Ei}(-2r\Lambda) - \log (2r\Lambda) - \gamma \right),
\end{align}\end{subequations}
where $\gamma\approx 0.577$ is the Euler-Mascheroni constant.   Note that $A(r)$ is only weakly modified at this order, while $B(r)$ has a logarithmic divergence.  The proposal of \cite{santosdisorder1} is that this logarithmic divergence is resummed at $r\Lambda \gg 1$ to create a Lifshitz geometry of the form \cite{kachru}
\begin{subequations}\label{eq:lifshitzgeometry}\begin{align}
A(r) &= \frac{A_0}{r^2}, \\
B(r) &= \frac{B_0 \Lambda^{2(1-z)}}{r^{2z}},
\end{align}\end{subequations}
where $A_0$ and $B_0$ are positive dimensionless constants, and $z$ is given by (\ref{eq:zlifshitz}). 
Such a geometry is dual to a field theory which is scale invariant but where time and space scale differently.   

The purpose of this paper is to analyze this conclusion more carefully when $\bar V \ll 1$.  We will find, via a heuristic but manifestly non-perturbative analysis, that $A_0, B_0 \sim 1$ are well-approximated by simple constants, while for $r\Lambda \gg 1$ \begin{equation}
z(r) \sim 1+ \frac{1}{\log (r\Lambda)} \log \left(1+\frac{\bar V^2}{8}\log (r\Lambda)\right).  \label{eq:zrdependence}
\end{equation}
While $\bar V \ll 1$ is treated as a small parameter throughout the analysis, we emphasize that (\ref{eq:zrdependence}) is a non-perturbative result because it extends to arbitrarily large values of $r$.  Indeed, (\ref{eq:zrdependence}) implies the following: there is an emergent infrared energy scale \begin{equation}
\Lambda_* \sim \Lambda \mathrm{e}^{-c/\bar V^2} \label{eq:Lambdastar}
\end{equation}
for $c$ an O(1) constant which we might estimate as $c\approx 8$.  At energy scales $E$ obeying $\Lambda_* \ll E \ll \Lambda$, the theory is approximately Lifshitz scale invariant with dynamical critical exponent (\ref{eq:zlifshitz}).   However, at energy scales $E \ll \Lambda_*$, the Lifshitz scaling breaks down.  We argue that the previous analysis in \cite{santosdisorder1} is only valid for energy scales $E\gtrsim \Lambda_*$, where their results match ours.  But we claim that the true infrared theory is, upon rescaling $t$ by a $\bar V$-dependent factor, dual to a holographic model with the following unusual metric: \begin{equation}
\mathrm{d}s^2 \sim \frac{\mathrm{d}r^2+\mathrm{d}x^2}{r^2} - \frac{\mathrm{d}t^2}{r^2 \log^2 (r\Lambda)}. \label{eq:trueIR}
\end{equation}
This theory is almost -- but not quite -- scale invariant in the infrared.  The proposal that (\ref{eq:trueIR}) describes the genuine infrared physics is the main result of our paper. 

Another appearance of logarithmically running couplings in a holographic model arises in Einsten-Maxwell-dilaton models \cite{mybook}, where a scalar field $\phi \sim \log r$ grows logarithmically in the IR.  However, the metric supported by the dilaton (and a gauge field) does not have an explicit logarithm as in (\ref{eq:trueIR}).  For this reason, we expect that (\ref{eq:trueIR}) represents a subtly different and new kind of IR theory.  Further work exploring the response functions of such a geometry is worthwhile, and it would be interesting if the logarithm leads to slow dynamics reminiscent of aging or glassiness.

Let us now justify (\ref{eq:zrdependence}).  We imagine that the ansatz (\ref{eq:lifshitzgeometry}) is correct, but that once $r\Lambda \gg 1$, $A_0$, $B_0$ and $z$ are slowly varying functions of $r$, with $A_0,B_0\sim 1$ and $z-1\ll 1$.  These assumptions are all consistent with the picture above.  We now show that a self-consistent solution to the holographic equations of motion with these properties exists.  

The first step is to evaluate the disorder averages over the scalar field profiles.  There are two key observations.  Firstly, since the scalar equation of motion \begin{equation}
\frac{r}{\sqrt{AB}} \partial_r \left(\sqrt{AB} r \partial_r \phi \right)= \frac{k^2}{A}\phi -\frac{3}{4}\phi \label{eq:scalarEOM}
\end{equation}
is linear, we may solve it separately for each $k$, temporarily assuming that we know $A_0$, $B_0$ and $z$ (we will self-consistently determine these later).   The second observation is that for $k\ll \Lambda$, if the geometry varies sufficiently slowly, the solution of  (\ref{eq:scalarEOM}) is well approximated by the $k=0$ solution for $r \ll k^{-1}$, and exponentially decays for $r\gg k^{-1}$.   To be precise, let us suppose that $A_0$, $B_0$ and $z$ from (\ref{eq:lifshitzgeometry}) are constants.  In this case, we can analytically solve (\ref{eq:scalarEOM}), up to a dimensionless prefactor $C\approx 1$ which will not play an important role in the analysis: 
\begin{equation}
\phi(k,r) \approx C \times h(k) \times \Lambda^{(1-z)/2} k^{z-1/2} r^{(1+z)/2}\mathrm{K}_{z-1/2}(\sqrt{A_0}kr) \label{eq:phiz}
\end{equation}
Note that the $\Lambda$ dependence arises in this equation because for $r\Lambda \lesssim 1$, the metric transitions to an asymptotically AdS region where $z=1$, and that since the metric is always close to $z=1$, and $C=1$ is the exact prefactor at $z=1$, it is reasonable to estimate $C\approx 1$.   Upon integrating over $k$, for $r\Lambda \gg 1$ we find the scaling relation \begin{equation}
\mathbb{E}\left[\phi^2\right] \sim r^2 \mathbb{E}\left[(\partial_x\phi)^2\right] \sim r^2 \mathbb{E}\left[(\partial_r\phi)^2\right] \sim \bar V^2 (r\Lambda)^{1-z}. \label{eq:scalarsourceE}
\end{equation}
We emphasize that there is now $z$-dependence in these equations, which will qualitatively change the infrared geometry.  This effect  was missing in the resummed perturbative analysis of \cite{santosdisorder1}.

In fact, $z$ as given in (\ref{eq:zrdependence}) is not strictly constant.  Nevertheless, it is reasonable to approximate that it is constant in the solution (\ref{eq:phiz}).  Firstly, let us ask whether the $r$-dependence of $z$ can affect the exponential fall-off once $kr\gg 1$ and $\Lambda_*r \gg 1$: from (\ref{eq:trueIR}) we estimate \begin{equation}
r\frac{\mathrm{d} z}{\mathrm{d}r} \sim \frac{\log\log(r\Lambda)}{\log^2 (r\Lambda) } \ll 1,
\end{equation} 
and therefore on the scales where any given wave number scalar mode are decaying, the metric will approximately appear scale invariant (in the Lifshitz regime we also have approximate scale invariance).  Secondly, for $kr\ll 1$, it is reasonable to approximate $\phi(k,r)$ with the solution at $k=0$.  At $k=0$, it is straightforward to use the method of dominant balance \cite{bender} to check that (\ref{eq:phiz}) is a very good approximate solution to the scalar equation of motion with an effective $z(r)$ given by (\ref{eq:zrdependence}).

Having justified that in the deep IR $\Lambda_*r \gtrsim 1$, we may approximate (\ref{eq:scalarsourceE}), we now must confirm that indeed (\ref{eq:zrdependence}) is a self-consistent solution to the Einstein equations.  Again we use the method of dominant balance.  First, we note that a solution with $z$ constant (and $A_0$, $B_0$ constant) cannot possibly solve (\ref{eq:system}) even approximately: the Lifshitz geometry is supported by $\mathbb{E}[\phi^2]\sim (r\Lambda)^0$, and yet in the Lifshitz regime we instead find (\ref{eq:scalarsourceE}).  Since $z>1$ was supported by the scalar disorder, the most natural guess as to how to correct the solution is to assume $A_0$ and $B_0$ remain constants of order 1, while $z(r)$ is weakly dependent on $r$.  After all, we expect that $\mathrm{d}z/\mathrm{d}r<0$ in the IR (at least at $r\sim \Lambda_*^{-1}$) due to the irrelevance of the disorder in the Lifshitz regime.  Upon plugging in the ansatz (\ref{eq:lifshitzgeometry}), with $z(r)$ an unknown function, into (\ref{eq:system}), we find the approximate equation \begin{equation}
\frac{6(z-1)}{r^2} + \frac{4}{r}(\log(r\Lambda)-1)\frac{\mathrm{d}z}{\mathrm{d}r} - 2\log(r\Lambda)\frac{\mathrm{d}^2z}{\mathrm{d}r^2}  \approx \frac{3C}{4r^2} \bar V^2 (r\Lambda)^{1-z}
\end{equation}
where we have dropped terms which are higher order in $z-1$.  We expect that $C\approx 1$ for the same reasons as before.  When $C=1$, we can use dominant balance to show that (\ref{eq:zrdependence}) is an approximate solution to this equation.  This demonstrates that as promised, (\ref{eq:zrdependence}) is a self-consistent approximate solution to the holographic equations, in our ``mean field" approximation for disorder.   

A non-trivial feature of the infrared solution we have found is the apparent independence of (\ref{eq:trueIR}) on $\bar V$ -- even though the disordered scalar is still supporting the geometry.  What our solution suggests is the following: combining (\ref{eq:zrdependence}) with (\ref{eq:scalarsourceE}), we obtain \begin{equation}
\mathbb{E}\left[\phi^2\right] \sim \frac{\bar V^2}{1+ \frac{1}{8}\bar V^2 \log (\Lambda r)}.  \label{eq:Ephi2}
\end{equation}
The disorder strength $\bar V$ drops out of this expression to leading order when $\Lambda_* r\gg 1$ because the effective dynamical critical exponent $z$, which itself depends on the disorder strength, reduces the bulk scalar field's magnitude in the infrared.   It is tempting to thus view the scalar as a dangerous ``marginally irrelevant" operator in the true infrared theory;  however, since the infrared theory is not scale invariant, we are not sure if the standard terminology fully captures this physics.  

While on the subject of relevant versus irrelevant perturbations, however, we call attention to the following simple observation.   Suppose that our conjecture is wrong, and that the genuine infrared theory at all scales is Lifshitz, as proposed in \cite{santosdisorder1}.  Then, one might expect that the disorder which drove us to the Lifshitz fixed point itself becomes irrelevant.  But that does not appear quite true -- we have a line of fixed points which depend on $\bar V$, and so the disorder remains a ``relevant" operator even in the Lifshitz geometry.  However, in our proposed infrared description, modifying the disorder strength does appear irrelevant, in the sense that (\ref{eq:Ephi2}) does not depend on $\bar V^2$ at leading order at large $r$.  Hence our proposed infrared theory does have a desired property that the operator which perturbed us to this fixed point at high energies has actually become ``irrelevant" at low energies.

The last thing to check is that $A_0$ and $B_0$ do not significantly depend on $r$.  In some sense, $B_0$ can exactly be set to a constant as all the $r$-dependence in $B(r)$ can be absorbed into $z(r)$.  To check that $A_0$ is well approximated by a constant, we again apply the principle of dominant balance.  Plugging in the ansatz (\ref{eq:lifshitzgeometry}), (\ref{eq:zrdependence}) and (\ref{eq:scalarsourceE}) into (\ref{eq:system}), we indeed find that the error terms in the differential equations are smaller than the leading order terms, and source subleading corrections to $A_0(r)$.  This completes our self-consistency check.

We emphasize that it is not the case that keeping higher order terms in $\bar V^2$ can restore the Lifshitz geometry.  As noted before in (\ref{eq:deltalifshitz}), so long as the geometry is approximately homogeneous, it will ultimately drive the scalar disorder to irrelevance.  No $\bar V^4$ corrections to any coefficients above will change that physics, so long as $\bar V$ is sufficiently small.

\section{Nonlinear stabilization of Lifshitz geometry}
Thus far, we have described a minimal model of disordered holography.  We now turn to a discussion of a more general family of theories, with bulk action  \begin{equation}
S = \int \mathrm{d}^3x \sqrt{-g}\left(R + 2 - \frac{1}{2}\nabla_a \phi \nabla^a \phi - \frac{1}{2}m^2 \phi^2 - \frac{1}{4}g\phi^4 - \cdots \right), \label{eq:sec4S}
\end{equation}
where for convenience we have only included even terms in $\phi$.  As before, we choose $m^2=-3/4$ so that the disorder is Harris-marginal.  While the term above will modify (\ref{eq:system}), it is a subleading effect $\sim \bar V^4$ and its effect on the geometry is (for the moment) not  important.  However, the quartic term does modify the scalar equation of motion in an important way.   In the same spirit as our ``mean field" approximation (\ref{eq:system}), the scalar equation of motion (\ref{eq:scalarEOM}) generalizes to \begin{equation}
\frac{r}{\sqrt{AB}} \partial_r \left(\sqrt{AB} r \partial_r \phi \right)= \frac{k^2}{A}\phi -\frac{3}{4}\phi +3g\mathbb{E}\left[\phi^2\right] \phi . \label{eq:gscalar}
\end{equation}
Due to the nonlinearity, once $r\Lambda \gg 1$, the scalar field has an effective mass \begin{equation}
m_{\mathrm{eff}}^2 = -\frac{3}{4} + 3g\mathbb{E}\left[\phi^2\right].  \label{eq:meff}
\end{equation}
In the Lifshitz regime $\Lambda^{-1} \ll r \ll \Lambda_*^{-1}$, we can evaluate the criterion for Harris-marginal disorder with dynamical critical exponent $z$: combining (\ref{eq:zlifshitz}) and (\ref{eq:harrismarginal}), we require \begin{equation}
\Delta_{\mathrm{eff}} = \frac{3}{2} + \frac{\bar V^2}{8} + \cdots . \label{eq:deltaeff}
\end{equation}
Combining (\ref{eq:mdelta}), (\ref{eq:meff}) and (\ref{eq:deltaeff}), we see that when \begin{equation}
g = -\frac{1}{48} + \cdots,
\end{equation}
the disorder remains marginal in the Lifshitz regime, at leading non-trivial order in $\bar V$.

The solution to (\ref{eq:gscalar}) with this value of $g$ is approximately given by \begin{equation}
\phi(k,r) \sim h(k) \times k^{(1/2+\bar V^2/16)} r^{1+\bar V^2/16} \mathrm{K}_{1/2+\bar V^2/16}(kr).
\end{equation}
Hence, at leading order in $\bar V$, we find that \begin{equation}
\mathbb{E}\left[\phi^2\right] = \bar V^2 = 4\; \mathbb{E}\left[(\partial_x\phi)^2 - (\partial_r\phi)^2\right].  \label{eq:ephi2}
\end{equation}
Plugging (\ref{eq:ephi2}) into (\ref{eq:system}), we find that the Lifshitz geometry (\ref{eq:lifshitzgeometry}) with \emph{constant} $z$ given by (\ref{eq:zlifshitz}) is a good solution.   

However, since we have neglected $\bar V^4$ corrections both in the value of $z$, and by neglecting the $\mathbb{E}\left[\phi^4\right]$ term in system, we should not expect that for \emph{all} $r$ is the geometry Lifshitz.    Rather, we expect that at a very infrared scale energy scale of $\Lambda_* \sim \Lambda \exp[-1/\bar V^4]$, a new infrared theory emerges.   In principle that regime could in turn be stabilized by higher order nonlinearites, etc., but we expect that a true Lifshitz geometry only emerges for an infinitely fine tuned scalar potential $V(\phi)$ in the holographic action.  Of course, since the coefficients such as $m^2$, $g$, etc. are related to operator dimensions and operator product expansion coefficients by the holographic dictionary, such fine tuning may be rather unreasonable from a field theoretic point of view.

The presence of odd terms in $\phi$ in (\ref{eq:sec4S}), which are actually quite natural in holography, seems to have a very drastic effect.  Odd terms in $\phi$ will generate a non-trivial $k=0$ component to the scalar field in the presence of disorder, since there is now a $\phi^2$ term in (\ref{eq:gscalar}).  This may lead to a very different infrared theory.   We will not perform a detailed analysis here.   We also note that in any finite size realization of a holographic theory whose action is invariant under $\phi \rightarrow -\phi$, in principle $k=0$ components can also arise due to an interplay between the metric components at wave number $q$ and scalar field at wave number $-q$; however, in the thermodynamic limit the theory could not pick how the zero mode $\phi$ breaks this $\mathbb{Z}_2$ symmetry in the absence of a source, and so we expect our analysis neglecting zero modes of $\phi$ is acceptable for all theories studied in this paper.

\section{Speculation on relevant disorder}
For $g>-\frac{1}{48}$, the nonlinearity above will not drastically change the physics relative to $g=0$.  However, if $g<-\frac{1}{48}$, a qualitatively new infrared theory can arise.  

We conjecture that the low energy physics of the regime $g<-\frac{1}{48}$ is qualitatively similar to a theory with Harris-relevant disorder.  We assume again that the disorder is given by (\ref{eq:momspacedis}), although we may safely now take $\Lambda\rightarrow\infty$.    In our one dimensional model, this corresponds to $\Delta<\frac{3}{2}$.    Assuming $\Delta>1$ for convenience, one can solve the equations of motion and find \cite{mybook} \begin{equation}
\mathbb{E}\left[\phi^2\right] \sim r^2 \mathbb{E}\left[(\partial_x\phi)^2\right] \sim r^2 \mathbb{E}\left[(\partial_r\phi)^2\right] \sim \bar V^2 r^{3-2\Delta}. \label{eq:relevantsource}
\end{equation}
The infrared energy scale is now simply given by \begin{equation}
\Lambda_* \sim \bar V^{1/(3/2-\Delta)}.
\end{equation}
For $\Lambda_* r \ll 1$, the theory is effectively conformal.  For $\Lambda_* r \gg 1$, we can crudely explore the infrared geometry as follows.  If we naively plug (\ref{eq:relevantsource}) into (\ref{eq:system}):  then we find using dominant balance that \begin{equation}
\log A(r) \sim \log B(r) \sim -\bar V r^{\frac{3}{2}-\Delta}
\end{equation}
once $\Lambda_* r \gg 1$.   This suggests that the geometry wants to ``cap off", as in a gapped phase \cite{mybook}.  

Unfortunately, the iterative process of correcting $\phi$, and subsequently $A$ and $B$, is badly divergent.  Therefore, we encourage numerical studies of the low energy holographic descriptions of such models.

\section{Conclusion}
In this paper, we have argued that Harris-marginal disorder does not drive the minimal holographic theory (\ref{eq:minimal}) to a Lifshitz scale invariant theory with a tunable dynamical critical exponent, as was proposed in \cite{santosdisorder1, aharony}.  Instead, we argue that at least for sufficiently small disorder, a universal infrared theory emerges below a non-perturbatively small energy scale.  The physics in the infrared appears independent of the disorder strength.   As our conclusion relies on manifestly non-perturbative physics, it is not surprising that these effects have not been seen in resummed perturbative analyses, which are appropriate at intermediate energy scales but need not capture all non-perturbative physics.   

As far as we can tell from earlier numerics \cite{santosdisorder1}, our assumption that the essential physics of the geometry is captured by a homogeneous scaling geometry of the form (\ref{eq:homogeometry}) seems to be the case: there is a well-defined Lifshitz exponent (at numerically accessible scales) which is quantitatively predicted by this approach \cite{mybook}.  So we do expect that our approach ultimately captures the essential infrared physics and that our qualitative conclusion is correct.  

Due to the possibly extraordinarily small energy scales at which the true infrared regime sets in, it may well be impossible to observe this effect in a numerical simulation.  Let us make a brief estimate.  Assuming (\ref{eq:Lambdastar}) is accurate for all $\bar V$, the numerics of \cite{santosdisorder1} reached $\Lambda r \sim 10^2$, meaning that when $\bar V \approx 1.3$, we expect them to just begin to see the onset of the true infrared.  By this estimate, the effect should be visible.  On the other hand, we can ask how much $z(r)$ as given in (\ref{eq:zrdependence}) would change between $10<\Lambda r < 100$;  again taking (\ref{eq:zrdependence}) at face value, we find that the answer is at most a few percent, which does not seem to contradict numerics.   We also emphasize, however, that these estimates are quite naive: (\ref{eq:zrdependence}) is certainly not accurate for $r\Lambda \sim 1$, and is genuinely intended only for the regme $\Lambda_*r \ll 1$.  Moreover, only when $\bar V \ll 1$ are the O(1) factors in any of our expressions believable, and this condition is certainly violated at $\bar V \approx 1.3$.   With all of these caveats, it is reasonable to conjecture that even when $\bar V > 1$, the infrared limit is the same as what we have obtained.  After all, our arguments around (\ref{eq:deltalifshitz}) do not rely on a perturbatively small $\bar V$.


We believe that our line of argumentation, while no doubt heuristic, is quite powerful and may be useful in the analysis of disordered holographic matter whenever the inhomogeneity arising from the disorder does not lead to ``rare region effects" where the inhomogeneity controls the physics.  Hence, it would be interesting to extend our study to other models of holographic disorder.  For example, our methods may lead to a more ``microscopic" derivation of existing ``mean-field" treatments of holographic disorder \cite{donos2, vegh, davison, blake, donos1, andrade, gouteraux, woodhead, pujolas}. Earlier work \cite{yaida} argued that chemical potential disorder (which couples to vector, not scalar, fields) is marginally relevant.  Whether this marginal relevance only builds up the Lifshitz regime \cite{awpeet, garciagarcia}, only to ultimately disappear in the true infrared, or whether this marginal relevance drives to a totally different ground state, is an interesting question for future research.

\section*{Acknowledgments}
\addcontentsline{toc}{section}{Acknowledgments}
We thank Mike Blake for useful feedback.

\begin{appendix}

\end{appendix}

\bibliographystyle{unsrt}
\addcontentsline{toc}{section}{References}
\bibliography{disorderbib}

\end{document}